\documentclass[11pt]{article}
\usepackage{fa2023}
\usepackage{amsmath}
\usepackage{cite}
\usepackage{url}
\usepackage{graphicx}
\usepackage{color}
\usepackage{siunitx}
\usepackage[utf8]{inputenc}

\title{Reverberant Sound Field Equalisation for an Enhanced Stereo Playback Experience}

\multauthor{James Brooks-Park$^{1*}$ \hspace{1cm} Steven van de Par$^1$ } { \\
  $^1$ Acoustics Group, Cluster of Excellence Hearing4All, CvO University Oldenburg, Germany\\
\correspondingauthor{James.Brooks-Park@uni-Oldenburg.de}{First author et al.}
}

\usetikzlibrary{calc}

\def\centerarc[#1](#2)(#3:#4:#5)
    { \draw[#1] ($(#2)+({#5*cos(#3)},{#5*sin(#3)})$) arc (#3:#4:#5); }

\def\Waves[#1](#2)(#3:#4);{
  \begin{scope}
    \centerarc[#1](#2)(#3:#4:0.2)
    \centerarc[#1](#2)(#3:#4:0.4)
    \centerarc[#1](#2)(#3:#4:0.6)
  \end{scope}
}

\usetikzlibrary{shapes.geometric}
\def\speaker(#1)(#2)(#3);{
    \begin{scope}[shift={(#1)},rotate={#2},scale={#3}]
        \draw[line width = .1mm] (-0.8,-0.40) rectangle (-0.35,0.40);
        \draw[line width = .1mm] (-0.35,0.40) -- (0,0.65) -- (0,-0.65) -- (-0.35,-0.40);
        \Waves[gray,thick](0,0)(60:0);
        \Waves[gray,thick](0,0)(360:300);
    \end{scope}
}
\def\Head(#1)(#2)(#3);{
    \begin{scope}[shift={(#1)},rotate={#2},scale={#3}]
      \draw[line width = .1mm](0,0) circle (1.5);
      \draw[line width = .1mm] (-0.15,1.5)--(0,1.8) -- (0.15,1.5);
      \draw[line width = .1mm] (-1.56,0) ellipse (0.1 and 0.2);
      \draw[line width = .1mm] (1.56,0) ellipse (0.1 and 0.2);
    \end{scope}
}

\def\RoomLayout(#1)(#2,#3);{
  \begin{scope}[scale={#1}]
    \scalefont{#1}
    \begin{tikzpicture}
      \draw[line width = .15mm] ( {(#2+0) *#1}, {(0+#3) *#1}) -- ( {(37+#2) *#1}, {(0+#3) *#1}) -- ( {(37+#2) *#1}, {(45+#3) *#1}) -- ( {(0+#2) *#1}, {(45+#3) *#1}) -- cycle;
      \speaker({(8+#2)*#1},{(38+#3)*#1})(310)(4*#1);
      \speaker({(29+#2)*#1},{(38+#3)*#1})(230)(4*#1);
      \speaker({(8+#2)*#1},{(7+#3)*#1})(45)(3*#1);
      \speaker({(29+#2)*#1},{(7+#3)*#1})(135)(3*#1);
      \draw[-latex,black, line width = .1mm] ({(8.4+#2)*#1},{(37.6+#3)*#1}) -- ({(15.5+#2)*#1},{(26+#3)*#1}) ;
      \draw[gray, line width=.1mm, -latex] ({(8.4+#2)*#1},{(7.6+#3)*#1}) -- ({(15.5+#2)*#1},{(24+#3)*#1});
      \Head({(18.5+#2)*#1},{(25+#3)*#1})(0)(1.5*#1);
    \end{tikzpicture}
  \end{scope}
}

\usepackage{scalefnt}
\usepackage{enumitem}
\setlist{itemsep=1pt, topsep=2pt}
\usepackage[moderate]{savetrees} 
\usepackage{float}
\begin{document}

\maketitle

\begin{abstract}
The topic of room equalisation has been at the forefront of research and product development for many years, with the aim of increasing the playback quality of loudspeakers in reverberant rooms. Traditional room equalisation systems comprise of a number of filters that when applied to the primary loudspeakers, additional room colouration is compensated for. This publication introduces a novel equalisation technique  where gammatone filter band energy is added to the reverberant sound field via two surround loudspeakers, leaving the direct sound from the primary loudspeakers unaltered, but the sum of direct  and reverberant energy is equalised at the listening position. Unlike traditional systems, this method allows the target function of the direct sound to differ from the reverberant sound field. The proposed method is motivated by the different roles direct and reverberant sound components play in humans' perception of sound. Along with introducing the proposed method, results from a subjective listening test are presented, demonstrating the preference towards the proposed technique when compared to a traditional room equalisation technique and stereo playback.
\end{abstract}

\vspace{-.3cm}
\section{Introduction}\label{sec:introduction}
When purchasing a traditional Hi-Fi system, one may spend days researching each component to create a system that perfectly matches ones acoustical and budgetary needs. When in fact, in a room with no acoustical treatment or equalisation, the perceived sound will be clouded by the acoustical colouration of the listening room. Therefore, room equalisation solutions may be highly desirable for those who do not wish to redesign their listening room to reach the full potential of their Hi-Fi system.

Traditional room-equalisation approaches typically aim to design several filters, so that when applied to the playback speakers, additional colouration caused by imperfect reflections in the playback room are compensated for~\cite{RRE}. Whilst methods that take this approach can quite effectively compensate for reverberant colouration in the playback room, the direct sound is inherently altered. The importance direct sound has on listeners perception of a loudspeakers spectral balance and timbre has been shown in~\cite{olive2004multiple}. Taking this into consideration, approaches have been developed that aim to compensate for direct and reverberant sound independently. The approach presented in~\cite{Quasi2013} uses separate equalisation methods for the direct and reverberant sound, using a non-minimum-phase FIR filter for the quasi-anechoic response and a fractional octave resolution minimum phase equaliser to equalise the total response. However, as the filters are applied to the primary loudspeakers, the direct sound is also altered.

Previous literature shows, that when room equalisation target functions are under development, it is often the case that a downward spectral slope towards the higher frequencies is preferred\cite{olive2013listener,olive2009subjective,toole2017sound}. One could speculate that this is because a natural, in-room transfer function of a conventional loudspeaker would also follow this trend. Therefore, spectrally flat target functions may sound harsh and unnatural if used for total sound field equalisation.

Whilst it is important to control the additional colouration introduced by the playback room, it would be undesirable to remove all reverberant energy from the listening experience. The effect reverberation has on sound reproduction is well documented. In~\cite{kaplanis2019perception}, the role reverberation has on the perception of loudspeakers in small listening rooms is investigated, showing its effect on the perceived timbre and spatial characteristics. Therefore for perceptually motivated room equalisation methods,it may be desirable to optimize reverberation with regards to it effect on timbre and spatial characteristics. The work presented in~\cite{grosse2015}, reproduces both the recorded audio and the recording room acoustics in the playback room, by recording the direct and reverberant sound independently, they can then be separately reproduced and optimised in the playback room.

The novel method proposed in this publication assumes that the on-axis frequency response of the primary loudspeakers is well optimised by the manufacturer, however the combination of speaker directivity and the listening room affect the total in-room response. The proposed approach aims to compensate for the reverberant sound field colouration, by altering the reverberant sound field using supporting surround loudspeakers with a gammatone filter bank based equaliser, in order to achieve total sound field equalisation.

With the increasing popularity of distributed surround loudspeakers in the home, and the extensive library of stereo recordings, the proposed approach aims to take advantage of modern audio reproduction systems to increase the listening experience of a classical media format. Along with presenting the proposed method in Section~\ref{sec:Method}, results from a subjective listening experiment are discussed in Section~\ref{sec:Subjective}, validating the improvement the proposed approach achieves compared to conventional stereo playback.

\vspace{-.3cm}
\section{Proposed Method}\label{sec:Method}
The proposed method aims to alter the reverberant sound  field,to compensate for additional room colouration. In order to compensate for playback room colouration, acoustical information about the room and reproduction system is captured by taking room impulse response measurements of the primary loudspeakers at the listening position. Two measurement microphones are positioned 17cm apart and an average of the two impulse responses is calculated for each of the primary loudspeakers. To achieve equalisation at the listening position, energy is added to the reverberant sound field using supporting loudspeakers to reach a specified spectral target function. In accordance with previous literature, a spectrally sloping target function with a drop of 5dB across the audible range is used for the reverberant sound field compensation.

\begin{figure}
  \includegraphics*[width = \linewidth]{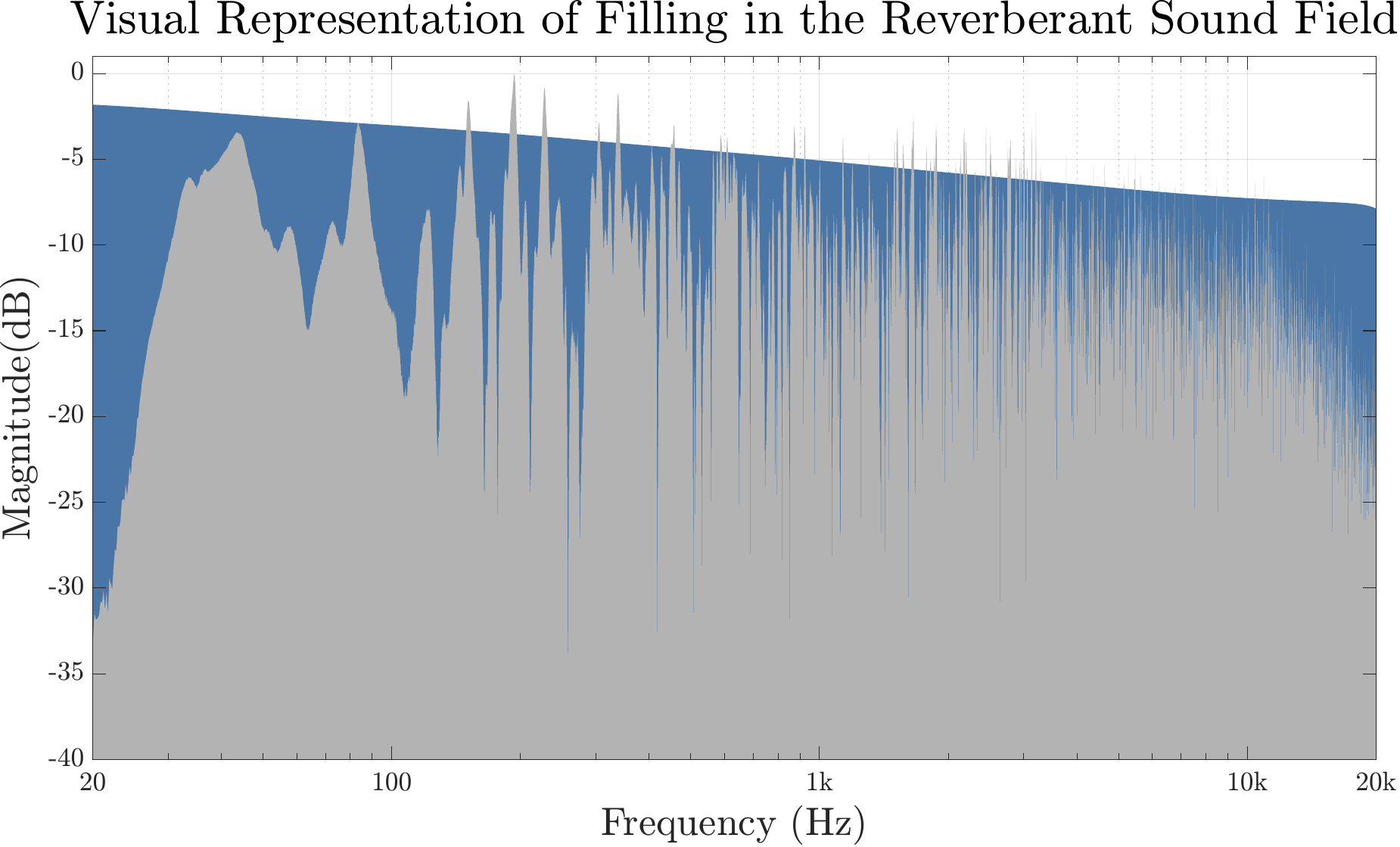}
  \vspace{-0.9cm}
  \caption[RIR]{\footnotesize Graphical representation of spectrally `filling in' acoustical energy with supporting loudspeakers to equalise the total sound field at the primary listening position. Grey = primary loudspeaker spectral response, Blue = filling in area}\label{RIR}
  \vspace{-0.6cm}
\end{figure}

To achieve total sound field equalisation at the listening position, scaling factors are calculated for a gammatone filter bank based equaliser,  so that when applied to the supporting loudspeakers the summation of the direct and reverberant energy matches that of the spectral target function. These scaling factors are calculated independently for the left and right loudspeaker and applied to their corresponding supporting loudspeaker. This proposed concept of filling in the missing spectral energy is visually presented in Figure~\ref{RIR}. Where the grey area represents the frequency domain transfer function of the primary loudspeakers, and the blue area represents the spectral energy the supporting loudspeakers aim to fill. The method presented in~\cite{van2005scalable}, and used in~\cite{grosse2015}, has been implemented to calculate scaling factors for overlapping gammatone filter bands, using the gammatone filter band analyser and synthesiser presented in \cite{hohmann2002frequency}. In order to accurately control the sound field at the primary listening position, the level of all loudspeakers are balanced to achieve equal SPL's at the  primary listening position.

To ensure the original stereo image of the reproduced stereo content is maintained, the supporting loudspeakers are delayed by 10ms. This ensures that the supporting loudspeakers are not perceived by the listener as the precedence effect comes into effect. The precedence effect states that, in a scenario with two sound sources, if one of the sources is delayed between 2-50ms, humans will perceive the source from the direction of the leading source. A 10ms delay has been deemed suitable for maintaining the original stereo image whilst avoiding echo artefacts in the reproduction system. To further ensure that the supporting loudspeakers become perceptually part of the reverberant sound field, an all-pass decorrelation filter from~\cite{kermit} is applied to the supporting loudspeakers. Decorrelating loudspeakers is a common technique for multichannel systems to ensure no phantom sources are created between loudspeakers and to remove any chance of comb filtering occurring. Both the delay and decorrelation are vital in ensuring the spatial imaging of playback media does not experience unwanted alterations.


\vspace{-.3cm}
\section{Listening Test}\label{sec:Subjective}
\subsection[Method and Subjects]{Method and Subjects}
A paired comparison listening test format was used to compare the proposed method against a number of test stimuli representing base line systems. Based upon the subjective rating of 4 attributes, participants were asked to rate the proposed method against a test stimulus using a 5 point scale. A score of 2 on this scale indicates a clear preference towards the proposed approach, 1 would indicate a just noticeable preference towards the proposed approach and 0 indicates no preference towards either stimuli. The opposing scores -1 and -2 indicate a preference towards the test stimulus. The test GUI allowed the subjects to switch between the test stimuli in real time, with no limit to the length of time they could listen to any one stimulus. The subjects  were presented with all attributes at the same time, once the subject moved on from the current test stimulus, they were unable to make changes to previous responses. The selected attributes and their definitions are as follows.

\noindent
\textbf{Preference:} The subjects overall subjective opinion of the presented stimuli, simply conveying which approach they would prefer to listen to given the material using.

\noindent
\textbf{Immersiveness:} Immersiveness is defined in this experiment as how psychologically connected to the stimuli the subject feels.

\noindent
\textbf{Clarity:} For any room equalisation approach the overall goal is to improve the clarity of an reproduction system. Subjects were asked to rate the stimuli based on both spectral and spatial clarity.

\noindent
\textbf{Naturalness:} This attribute allows the subject to rate how natural the presented stimuli would feel if they were to imagine the performer to be in the same room as them.

The primary loudspeakers used for this study were Kef LS50's and Genelec 6010A Monitors for the supporting loudspeakers. These were both connected to the Max 8 test software using an RME AD-8 QS D/A converter.

In order to select subjects for the listening test, 10 colleagues of the authors research group with no prior knowledge of the proposed method were asked to participate. All participants have experience with audio systems from a number of different backgrounds, ranging from signal processing research to a composer and mixing engineers. Prior to the main comparison based listening test, subjects were asked to participate in a short ABX test that presented the proposed method and the rear stereo stimuli (Detailed below). This short training phase allowed the participants to experience the stimulus before the main experiment and also allowed for a screening process to be undertaken. If the subject was not able to distinguish the difference between the stimuli, there responses would not be included in the results for this experiment.

\vspace{-.3cm}
\subsection[Stimy]{Stimuli}
Whilst the main aim of this study is to compare the proposed approach to stereo playback, a number of additional base-line reproduction methods have been selected to understand and isolate areas where the proposed approach may show improvement. The comparison methods were applied to 4 audio excerpts allowing subjects to rate each method applied to several genres of music.

\noindent
\textbf{Rear Stereo:}
The original stereo material is presented both through the primary loudspeakers and the rear speakers with no alterations.

\noindent
\textbf{Front Equalisation:}
Scaling factors are only applied to the front loudspeakers to make the total response follow the spectral target function.

\noindent 
\textbf{Stereo:}
Traditional stereo loudspeaker reproduction.

\subsection[ResSec]{Results}
\begin{figure}[H]
    \input{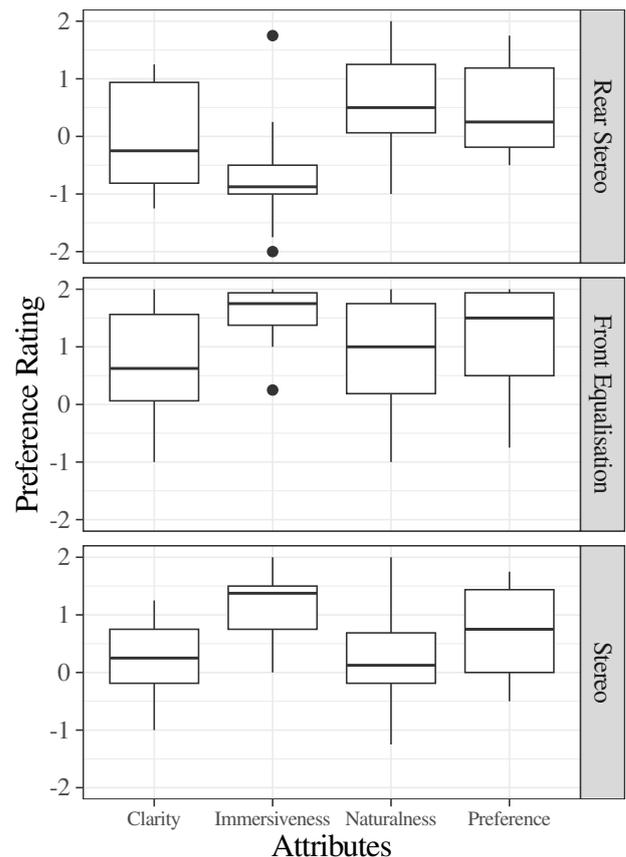}
    \vspace{-1cm}
    \caption[Results]{\footnotesize Preference study results comparing the proposed approach to three test stimuli across 4 attributes. A score of 2 indicates a clear
    preference for the proposed approach, 1 indicates a just noticeable
    preference towards the proposed approach and 0 indicates no preference. Opposing scores indicate a preference towards the test stimuli indicated in the grey box to the right of the plot.}\label{Results}
    \vspace{-.5cm}
\end{figure}

The results from this study can be observed in Figure~\ref{Results}. The plots show the median value for the subjects responses, the interquartile range (IQR), maximum and minimum values and the outliers that are greater than \begin{math}1.5 * IQR \end{math}.\newline
As can be seen from the comparison between the proposed method and the rear stereo stimuli (top panel), there is an increased level of immersiveness in the test stimuli. This is not consistent with the overall preference rating,  where there is a slight preference towards the proposed method. When looking at the results for the front equalisation comparison, it is clear that there is a trend showing preference towards the proposed method. The results for the stereo comparison show a small preference towards the proposed approach, with stronger preferences with regards to immersiveness.

\vspace{-.3cm}
\section{Discussion}\label{sec:Discussion}
The attributes used in the presented listening test  aim to isolate the areas of which the proposed approach shows improvement. The following section analyses the results from the study, to propose ideas as to how the proposed method increase the stereo in-room playback experience.

For the  ``rear stereo'' stimuli, the increased level of immersiveness is to be expected, due to an increased number of perceived sources. However, whilst the level of immersiveness is increased, the proposed approach maintains a noticeable preference.

Comparing the proposed method to the front equalisation stimulus, it can be seen that the proposed method shows better clarity results. This confirms  that the method of equalising the total sound field by altering the reverberant sound field is able to match and outperform a traditional approach where the direct sound is altered.

As the proposed method aims to improve playback of stereo media, the comparison between stereo playback and the proposed approach is perhaps the most important. As the discussed listening test took less than an hour to run, the results may differ if subjects were able to listen to the test stimuli over a longer period of time. It was mentioned by several subjects that their preference towards the constant stimuli (The proposed method) increased throughout the test. given this possibility,there remains a slight preference towards the proposed approach.


An advantage of the proposed method is compensation only occurs in the reverberant sound field, without altering the response of the primary loudspeakers. Meaning, the quality of the sound outside of the sweet spot is not degraded to the same level as with direct sound optimisation that aims to reduce the effect of the reverberant sound field. This is the case provided the listener is not so close to the supporting loudspeakers that the precedence effect is broken.

\vspace{-.3cm}
\section{Conclusion}\label{Conclusion}
This publication presents a novel room equalisation and stereo enhancement approach, aiming to compensate for additional room colouration at the listening position without altering the direct sound of the playback system. 

From the presented listening test, the following conclusions can be made. The preference towards the proposed approach is not simply due to the increased number of loudspeakers, although this may increase the level of immersiveness, the comparison to the rear stereo stimuli maintains an overall preference towards the proposed method. It can also be seen that the idea of equalising the sound field can be achieved by altering the reverberant sound field, differing from traditional methods whereby the direct sound field is subject to compensation. These results show that the presented approach enhances the listening experience of traditional stereo media across a range of attributes.

\vspace{-.3cm}
\section{Acknowledgments}
This project has received funding from the European Union's Horizon 2020 research and innovation programme under the Marie Skłodowska-Curie grant agreement No 956369.

\vspace{-.2cm}
 \bibliography{References.bib}

%
%
%

\end{document}